\documentclass[prl,aps,twocolumn, showpacs]{revtex4}
\usepackage[dvips]{graphicx}
\usepackage{dcolumn}
\usepackage[english, russian]{babel} %- для рус-англ. переноса
\usepackage{amsmath, latexsym,  amssymb, array, graphics,
amsfonts, amsthm, amsmath,  amsfonts, array, bm, eucal}

{

\begin{document}
\newcommand{\mc}[1]{\mathcal{#1}}
\newcommand{\E}{\mc{E}}
\topmargin=-15mm
%   \Large

\title {\bf
TEMPERATURE JUMP IN DEGENERATE QUANTUM GASES IN THE PRESENCE
OF A BOSE --- EINSTEIN CONDENSATE
}
%cond-mat/0601600

\author{\bf Anatoly V. Latyshev and Alexander Yushkanov}

\affiliation  {Department of Mathematical Analysis and
Department of Theoretical Physics, Moscow State Regional
University,  105005, Moscow, Radio st., 10--A}

\begin{abstract}
We construct a kinetic equation modeling the behavior of
degenerate quantum Bose gases whose collision rate depends
on the momentum of elementary excitations. We consider the
case where the phonon component is the decisive factor in
the elementary excitations. We analytically solve the half-space
boundary value problem of the temperature jump at the boundary of the
degenerate Bose gas in the presence of a Bose --- Einstein
condensate.\\

{\bf Keywords}: degenerate quantum Bose gas,
collision integral, Bose --- Einstein condensate, phonon component,
temperature jump, Kapitsa resistance.\\
\end{abstract}
%permeability - проницаемость
\pacs{05.20.Dd Kinetic theory, 05.30.Jp Boson systems,
05.60.-k Transport processes, 03.75.Nt Other Bose --- Einstein condensation
phenomena}%, 04.20.Jb Exact solutions}
\date{\today}
\maketitle

\section{1. Introduction}

The behavior of quantum gases has aroused increased interest in
recent years. In particular, this is related to the development of
experimental procedures for producing and studying quantum gases at
extremely low temperatures \cite{Pit}. The bulk properties of quantum gases
have been studied in the majority of papers
\cite{Spon} and \cite{Pang}.

At the same time, it is
obviously important to take boundary effects on the properties of such
systems into account. We mention a paper where the thermodynamic
equilibrium properties of quantum gases in a half--space were
considered \cite{Sam}.

Along with the equilibrium properties,
the nonequilibrium properties of quantum gases bounded in
space attract interest. In particular, such a phenomenon as
the temperature jump  \cite{Lan} at the interface between a gas and a
condensed (in particular, solid) body in the presence of a
heat flux normal to the surface is important. Such a temperature
jump is frequently called the Kapitsa
temperature jump \cite{Enz}.

We note that up to now, the Kapitsa jump has been calculated
in the regime where only phonon scattering at the interface
between two media was taken into account and phonon scattering
in the bulk was neglected \cite{H}. Here, we take namely the effect
of scattering of elementary excitations of the Bose gas (phonons)
in the bulk into account.

We take the character of phonon
scattering by the surface into account by introducing a
phenomenological coefficient of specular phonon scattering by
the surface. This approach is thus additional to that proposed
in \cite{H}.

In \cite{Lat1}, we considered the temperature jump problem in a quantum
Fermi gas. We obtained an analytic solution for an arbitrary
degree of gas degeneracy. In \cite{Lat2}, we considered a similar problem
for a Bose gas. But the gas was assumed to be nondegenerate,
i.e., we did not take the presence of a Bose --- Einstein condensate
into account.

This paper is devoted to analyzing the temperature jump problem
in a degenerate Bose gas. The presence of a Bose --- Einstein
condensate leads to a considerable modification of both
the problem statement and the solution method. In this case,
to describe kinetic processes near the surface, we use a kinetic
equation with a model collision integral. We assume that the
boundary conditions at the surface are specular--diffuse.

\section{2.  Derivation of the kinetic equation}

To describe the gas behavior, we use a kinetic equation with
a model collision integral analogous to that used to describe a
classical gas. We take the quantum character of the Bose gas and
the presence of the Bose --- Einstein condensate into account.
For a rarefied Bose gas, the evolution of the gas particle
distribution function $f$  can be described by the
kinetic equation \cite{Lan}
$$
\dfrac{\partial f}{\partial t} +
\dfrac{\partial \E}{\partial \mathbf{p}}
\nabla f=I[f],
\eqno{(1)}
$$
where $\E$ is the kinetic energy of gas particles, $\mathbf{p}$ is the gas
particle momentum, and $I[f]$ is the collision integral.

In the case of the kinetic description of a degenerate Bose gas,
we must take into account that the properties of the Bose --- Einstein
condensate can change as functions of the space and time coordinates,
i.e., we must consider a two--liquid model (more precisely, a
"two-fluid"\,
model because we consider a gas rather than a liquid).
We let $\rho_c=\rho_c({\mathbf{r}},t)$ and
${\mathbf{u}_c}={\mathbf{u}_c }({\mathbf{r}},t)$  denote the density and
velocity of the Bose --- Einstein condensate. We then can write the
expressions \cite{H}
$$
{\mathbf{j}}=\rho_c {\mathbf{u}_c},\qquad {\mathbf{Q}}=
\dfrac{\rho_c u_c^2}{2}{\mathbf{u}_c},
$$
$$
\Pi_{ik}=\rho_c u_{ci}u_{ck}.
$$
for the densities $\mathbf{j}$, ${\mathbf{Q}}$, and $\Pi_{ik}$
of the respective mass, energy,
and momentum fluxes of the Bose --- Einstein condensate
(under the assumption that the chemical potential is zero).
The conservation laws for the number of particles,
energy, and momentum require that the relations
$$
\nabla {\mathbf{j}}=
-\int I[f]d\Omega_B,
$$
$$
\nabla {\mathbf{} Q}=
-\int \varepsilon({\bf p}) I[f]d\Omega_B,
$$
$$
\nabla {\Pi}=
-\int {\mathbf{p}}I[f] d\Omega_B
$$
be satisfied in the stationary case. Here,
$$
d\Omega_B=\dfrac{(2s+1)d^3p}{(2\pi \hbar)^3},
$$
$s$ is the molecule spin, $\E({\bf p})$ is the energy, $\hbar$
is the Planck constant
and $I(f)$ is the collision integral in Eq. (1).

In what follows, we are interested in the case of motion with
small velocities (compared with thermal velocities). We note
that for the Bose --- Einstein condensate, the quantities ${\mathbf{Q}}$
and $\Pi_{ik}$
depend on the velocity nonlinearly (they are proportional to the
third and second powers of the velocity). Therefore, in the
approximation linear in the velocity ${\mathbf{u}_c}$, the
energy and momentum conservation laws can be written as
$$
\int \varepsilon({\bf p})I[f]d\Omega_B=0,
$$
$$
\int {\mathbf{p}}I[f]d\Omega_B=0.
$$

According to the Bogolyubov theory, the relation
for the excitation energy $\E(p)$ holds
for a weakly interacting Bose gas \cite{Lan}
$$
\varepsilon (p)=
\left[u_0^2p^2+\left(\dfrac{p^2}{2m}\right)^2\right]^{1/2},
\eqno{(2)}
$$
where
$$
u_0=\left(\dfrac{4\pi  \hbar^2an}{m^2}\right)^{1/2},
$$
$a$ is the scattering
length for gas molecules, $n$ is the concentration,
$m$ is the mass, and $\mathbf{p}$ is the momentum of elementary
excitations. The parameter $a$ characterizes the interaction
force of gas molecules and
can be assumed to be small for a weakly interacting gas.

In our previous paper \cite{Lat3}, we considered the case where
the relation
$$
u_0^2\ll \dfrac{kT}{m}
$$
holds for sufficiently small $a$,
where $k$ is the Boltzmann constant and $T$ is the temperature.
In that case, the first term in the brackets in (2) can be
neglected. The expression for the energy $\E(p)$ takes the same
form as in the case of noninteracting molecules:
$$
\E(p)=\dfrac{p^2}{2m}.
$$
We now consider the case where the phonon component dominates
in (2), i.e.,  where
$$
T\ll \dfrac{mu_0^2}{k}.
$$
In this case, we obtain
$$
\E(p)=u_0 |\mathbf{p}|=u_0p.% - \mathbf{p_0}|.
$$
according to relation (2). Consequently,
$$
\dfrac{\partial \E(\mathbf{p})}{\partial
\mathbf{p}}= u_0\dfrac{\mathbf{p}}{p}.
$$

When considering kinetic equation (1), by gas molecules,
we must understand the elementary excitations of the Bose
gas with energy spectrum (2). The character of the elementary
excitations is manifested in the properties of the collision
integral. As a collision integral in Eq. (1), we take its
$\tau$--approximation. Then the character of the elementary excitations
is manifested in the dependence
of the collision rate on the excitation momentum  \cite{Lat2},
\cite{Cerc}, \cite{Lat3}--\cite{Lat5}
$$
\dfrac{  \partial f}{ \partial t}+ u_0\dfrac{\mathbf{p}}{p}
\dfrac{\partial f}{\partial \mathbf{r}}=
\nu(\mathbf{p})(f^*_B-f).
\eqno{(3)}
$$

Here,  $f$  is the distribution function,
$$
\nu(\mathbf{p})=\nu_0|{\bf p}- {\bf p}_0|^\gamma
$$
is the dependence of the
collision rate on the excitation momentum,
and $\gamma$ is a constant. In the case where the phonon
component dominates in the elementary excitations,
$\gamma \geqslant 3$ \cite{Lan}. We have $\mathbf{p}_0=m{\bf v}_0$,
where  $\mathbf{v}_0$ is the velocity
of the normal component of the Bose gas, $f^*_B$ is the
equilibrium function of the Bose --- Einstein distribution
$$
f^*_B= \left[ \exp\left(\dfrac{u_0|\mathbf{p}-\mathbf{p_*|}}{kT_*}
\right)-1\right]^{-1},
$$
and $\nu_0$ is a model parameter having the
meaning of the inverse mean free path $l$, $\nu_0 \sim 1/l$.

The parameters in $f_B^*$/g, namely, $T_*$ and ${\bf u}_*$,
can be determined from the
requirement that the energy and momentum conservation laws
$$
\int \nu(\mathbf{p})\mathbf{p}\Big[f-f_B^*\Big]d^3p=0,
$$
$$
\int \nu(\mathbf{p})\varepsilon({\mathbf{p}})\Big[f-f_B^*\Big]d^3p=0
$$
be applicable. The conservation law for the number
of particles is inapplicable here because of the transition
of a fraction of particles to the Bose --- Einstein condensate.

We now assume that the gas velocity is much less than
the mean thermal velocity and the typical temperature
variations along the mean free path $l$ are small compared
with the gas temperature.
Under these assumptions, the problem can be linearized.

We seek the distribution function in the form
$$
f=f^{s}_B(C)+ \varphi(t,{\bf r},{\bf C})g(C),
$$
where
$$
f^{s}_B(C)=\dfrac{1}{\exp C-1},
$$
$\varphi$ is a new unknown function,
$T_s$ is the surface temperature,
$
g(v)=- \dfrac{
\partial }{ \partial \varepsilon_s}f^{s}_B,\quad
\varepsilon_s= C,
$
and we introduce the notation
$$
{\bf C}=\dfrac{u_0\,\mathbf{p}}{kT_s},
$$
and
$$
\varepsilon_*=\dfrac{u_0|\mathbf{p} - \mathbf{p_*|}}{kT_*}.
$$

We then have
$$
f^*_B(\varepsilon_*)= \dfrac{1}{\exp(\varepsilon_*)-1},
$$
$$
f^{s}_B(C)= \dfrac{1}{\exp C-1},
$$
$$
g(C)=\dfrac{\exp C}{(\exp C-1)^2}.
$$

We linearize the local Bose --- Einstein distribution (Bosean)
$f_B^*$, passing
to dimensionless quantities. We note that
$$
\varepsilon_*= \dfrac{T_s}{T_*}\cdot \dfrac{u_0|\mathbf{p} -
\mathbf{p_*}|}
{kT_s}= \dfrac{T_s}{T_s+\delta T_*}\cdot \dfrac{u_0}{kT_s}
\sqrt{(\mathbf{p}-\mathbf{p_*})^2}=
$$
$$
=\Big(1-\dfrac{\delta T_*}{T_s}\Big)\dfrac{u_0}{kT_s}\Big(p-
\dfrac{\mathbf{p} \mathbf{p_*}}{p}\Big),
$$
or
$$
\varepsilon_*=\Big(1-\dfrac{\delta T_*}{T_s}\Big)\Big(C-
\dfrac{\mathbf{C} \mathbf{C_*}}{C}\Big)=
C-\dfrac{\mathbf{C} \mathbf{C_*}}{C}-C
\dfrac{\delta T_*}{T_s}.
$$

Therefore
$$
\delta \varepsilon_*=\varepsilon_*-\varepsilon_s=-
\dfrac{\mathbf{C}\mathbf{C_*}}{C}-C\dfrac{\delta T_*}{T_s},
$$
because $\varepsilon_s=C.$

Consequently,
$$
f_B^*=f_B^s+g(C)\Big(\dfrac{\mathbf{C} \mathbf{C_*}}{C}+
C\dfrac{\delta T_*}{T_s}\Big).
$$

We note that the quantity $\nu(\mathbf{p})$ in Eq. (3) can be
replaced with $\nu_0p^\gamma$
in the approximation under consideration.

We introduce the dimensionless time
$$
\tau=\nu_0\Big(\dfrac{kT_s}{u_0}\Big)^\gamma t
$$
and coordinate
$$
{\bf r}_1=\nu_0\Big(\dfrac{kT_s}{u_0}\Big)^\gamma{\bf r}.
$$

It is now clear that
Eq. (3) (in the dimensionless variables) becomes
$$
\dfrac{\partial \varphi}{\partial \tau}+
\dfrac{{\bf C}}{C}\dfrac{\partial\varphi}{\partial \mathbf{r}_1}=
C^\gamma\Big[\dfrac{\mathbf{C} \mathbf{C_*}}{C}+
C\dfrac{\delta T_*}{T_s}-\varphi\Big].
\eqno{(4)}
$$

From the energy and momentum conservation laws, we find
$$
\dfrac{\delta T_*}{T_s}=\dfrac{1}{4\pi g_{\gamma+4}}
\int C^{\gamma+1}
\varphi(\tau,\mathbf{r}_1,\mathbf{C})g(C)d^3C,
$$
$$
\mathbf{} C_*=\dfrac{3}{4\pi g_{\gamma+3}}\int C^{\gamma}{\mathbf{C}}
\varphi(\tau,\mathbf{r}_1,\mathbf{C})g(C)d^3C,
$$
where
$$
g_{\gamma+n}=\int\limits_{0}^{\infty}C^{\gamma+n}g(C)\,dC,\qquad
n=0,1,2,\cdots.
$$

We represent Eq. (4) in the form that is standard in transport theory:
$$
\dfrac{\partial \varphi}{\partial \tau}+
\dfrac{\bf C}{C}\dfrac{\partial\varphi}{\partial \mathbf{r}_1}+
C^\gamma\varphi(\tau,{\bf r}_1,{\bf C})=
$$
$$
=
\dfrac{1}{4\pi}\int k({\bf C},{\bf C'})\varphi(\tau, {\bf r}_1,{\bf
C'}) g(C')d^3C,
\eqno{(5)}
$$
where $k(\mathbf{C},\mathbf{C'})$ is the kernel of Eq. (5),
$$
k(\mathbf{C},\mathbf{C'})=
\dfrac{3}{g_{\gamma+3}}C^{\gamma-1}{C'}^\gamma\mathbf{C}\mathbf{C'}+
\dfrac{1}{g_{\gamma+4}}C^{\gamma+1}{C'}^{\gamma+1}.
$$

Equation (5) can be represented in the equivalent form
$$
\dfrac{\partial \varphi}{\partial \tau}+
\dfrac{\bf C}{C}\dfrac{\partial\varphi}{\partial \mathbf{r}_1}+
C^\gamma\varphi(\tau,{\mathbf{r}_1},{\bf C})=
$$
$$
=\dfrac{3C^{\gamma-1}\mathbf{C}}{4\pi g_{\gamma+3}}
\int {C'}^\gamma\mathbf{C'}
\varphi g(C')d^3C'+
$$
$$
+\dfrac{C^{\gamma+1}}{4\pi g_{\gamma+4}}
\int {C'}^{\gamma+1}\varphi g(C')d^3C'.
$$

\section{3. Problem statement}

In the problem under consideration, a degenerate Bose
gas occupies the half--space  $x>0$ above a
planar surface where the heat exchange between
the condensed phase and the gas occurs. Therefore,
the function $\varphi$ can be regarded as
$$
\varphi(\tau, \mathbf{r}_1,\mathbf{C})=h(x,\mu,C)
$$
in what follows.
Such a function satisfies the equation
$$
\mu \dfrac{\partial h}{\partial
x}+C^\gamma h(x,\mu,C)=
$$
$$=
\dfrac{1}{2}\int\limits_{-1}^{1}\int\limits_{0}^{\infty}
K(\mu,C;\mu',C')h(x,\mu',C')g(C')d\mu' dC',
\eqno{(6)}
$$
where $K(\mu,C;\mu',C')$ is the kernel of Eq. (6),
$$
K(\mu,C;\mu',C')=\dfrac{3C^{\gamma-1}\mu
{C'}^{\gamma+3}\mu'}{g_{\gamma+3}}+
\dfrac{C^{\gamma+1}{C'}^{\gamma+3}}{g_{\gamma+4}}.
$$

The problem is to find the value of the relative
temperature jump
$$
\varepsilon_T=\dfrac{\Delta T}{T_s},
$$
where
$$
\Delta T=T_s-T,
$$
as a function of $Q_x$, which is the projection of the
heat flux on the $x$ axis. Taking linearity of the
problem into account, we can write
$$
\varepsilon_T=R Q_x.
$$

The dimensionless coefficient $R$ of the
temperature jump is called the Kapitsa resistance.

It is obvious that Eq. (6) has the particular
solutions
$$
h_1(x,\mu,C)=\mu\qquad\text{and}\qquad h_2(x,\mu,C)=C,
$$
and
the Chapman --- Enskog distribution function is
$$
h_{as}(x,\mu,C)=B^+\mu -\varepsilon_T C,
$$
where the quantity  $B^+$ is proportional to the heat flux $Q_x$.

Assuming that the reflection of the elementary
excitations from the wall is specular--diffuse,
we now formulate the boundary conditions
$$
h(0,\mu,C)=qh(0,-\mu,C),\quad 0< \mu<1,
\eqno{(7)}
$$
$$
h(x,\mu,C)=
$$
$$
=B^+\mu -\varepsilon_T C+o(1),\;
x\to +\infty,\;-1< \mu<0,
\eqno{(8)}
$$
where $q$ is the specular reflection coefficient.

The problem is to solve Eq. (6) with boundary conditions (7) and (8).
Finding the value of the temperature jump $\varepsilon_T$
is of special interest.

\section{4. Reduction to the integral equation}

We continue the function h to the half--space $x<0$ symmetrically:
$$
h(x,\mu,C)=h(-x,-\mu,C), \qquad x<0.
$$

For $x < 0$, we then have
$$
h_{as}(x,\mu,C)=B^-\mu-\varepsilon_TC,
$$
with $B^+=-B^-$.

 We now separate the Chapman --- Enskog distribution
 from the function $h$ assuming that
$$
h(x,\mu,C)=B^{\pm}\mu-\varepsilon_TC+h_c(x,\mu,C).
$$

For the function $h_c(x,\mu,C)$, we formulate the boundary conditions
for the lower and upper half--spaces:
$$
h_c(+0,\mu,C)=$$$$=-(1+q)B^+\mu+(1-q)\varepsilon_TC+q
h_c(+0,-\mu,C),
$$
where $\quad 0<\mu<1$,
$$
h_c(-0,\mu,C)=
$$
$$
=-(1+q)B^-\mu+(1-q)\varepsilon_TC+q
h_c(-0,-\mu,C),
$$
where $\quad -1<\mu<0$, and

$$
h_c(+\infty,\mu,C)=0, \qquad h_c(-\infty,\mu,C)=0.
$$

We include these boundary conditions in the kinetic equation.
We obtain the equation
$$
\dfrac{\mu}{C^\gamma}\dfrac{\partial h}{\partial x}+h_c(x,\mu,C)=
%$$$$=
\dfrac{C}{2g_{\gamma+4}}W_0(x)+\dfrac{3\mu}{2g_{\gamma+3}}W_1(x)+
$$$$+
|\mu|\{-(1+q)B^+|\mu|+$$$$+(1-q)\varepsilon_TC-(1-q)h_c(\mp 0,\mu,C)
\}\delta(x).
\eqno{(9)}
$$

Here, $\delta(x)$ is the Dirac delta function, and
$$
W_m(x)=$$$$+\int\limits_{-1}^{1}\int\limits_{0}^{\infty}
\mu^m C^{\gamma+3}h(x,\mu,C)g(C)d\mu dC,\quad m=0,1.
\eqno{(10)}
$$

Equation (9) actually combines two equations.
The point is that the term $h_c(-0,\mu,C)$
corresponds to positive $\mu: 0<\mu<1$
and the term $h_c(+0,\mu,C)$ corresponds to negative
$\mu: -1<\mu<0$.

We seek the solution of Eqs. (9) in the form of Fourier integrals:
$$
h_c(x,\mu,C)=\dfrac{1}{2\pi}\int\limits_{-\infty}^{\infty}
e^{ikx}\Phi(k,\mu,C)dk,
$$
$$
\delta(x)=\dfrac{1}{2\pi}\int\limits_{-\infty}^{\infty}
e^{ikx}dk,
$$
$$
W_0(x)=\dfrac{1}{2\pi}\int\limits_{-\infty}^{\infty}
e^{ikx}E_0(k)dk,
$$
$$
W_1(x)=\dfrac{1}{2\pi}\int\limits_{-\infty}^{\infty}
e^{ikx}E_1(k)dk.
$$

Solving Eq. (10) with $x > 0$ and $\mu<0$, assuming
that the right--hand side of this equation is known,
and assuming that the boundary
conditions are satisfied far from the wall, we obtain
$$
h_c^+(x,\mu,C)=$$$$=-\dfrac{C^\gamma}{\mu}
\exp\Big(-\dfrac{x}{\mu}C^\gamma\Big)\int\limits_{x}^{+\infty}
\exp\Big(\dfrac{t}{\mu}C^\gamma\Big)W(t,\mu,C)dt,
$$
where
$$
W(t,\mu,C)=\dfrac{C}{2g_{\gamma+4}}W_0(t,\mu,C)+
\dfrac{3\mu}{2g_{\gamma+3}}
W_1(t,\mu,C).
$$

After simple calculations, we obtain
$$
h_c^+(x,\mu,C)=\dfrac{C^\gamma}{2\pi}\int\limits_{-\infty}^{+\infty}
\dfrac{e^{ikx}E(k,\mu,C)dk}{C^\gamma+ik\mu},
$$
where
$$
E(k,\mu,C)=\dfrac{C}{2g_{\gamma+4}}E_0(k)+\dfrac{3\mu}{2g_{\gamma+3}}
E_1(k).
$$

We can similarly show that
$$
h_c^{-}(x,\mu,C)=\dfrac{C^{\gamma}}{2\pi}
\int\limits_{-\infty}^{+\infty}
\dfrac{e^{ikx}E(k,\mu,C)dk}{C^\gamma+ik\mu}.
%\eqno{(2.3)}
$$

From the last two expressions, taking the
evenness of the function $E(k,\mu,C)$
with respect to the variable $k$ into account, we obtain
$$
h_c^{\pm}(0,\mu,C)=$$$$=\dfrac{C^\gamma}{2\pi}
\int\limits_{-\infty}^{+\infty}\dfrac{(C^\gamma-ik\mu)E(k,\mu,C)dk}
{C^{2\gamma}+k^2\mu^2}=$$$$=\dfrac{C^{2\gamma}}{\pi}
\int\limits_{0}^{+\infty}\dfrac{E(k,\mu,C)dk}{C^{2\gamma}+k^2\mu^2}.
\eqno{(11)}
$$

Hence, we see that two Eqs. (9) can  be combined into one:
$$
\dfrac{\mu}{C^\gamma}\dfrac{\partial h}{\partial x}+h_c(x,\mu,C)=
$$$$=
\dfrac{C}{2g_{\gamma+4}}W_0(x)+\dfrac{3\mu}{2g_{\gamma+3}}W_1(x)+
$$$$+
|\mu|\{-(1+q)B^+|\mu|+$$$$+(1-q)\varepsilon_TC-(1-q)h_c^{\pm}(0,\mu,C)
\}\delta(x).
\eqno{(12)}
$$

We pass to the Fourier integrals in Eq. (12) and obtain the equation
$$
(C^\gamma+ik\mu)\Phi(k,\mu,C)=
\dfrac{C^{\gamma+1}}{2g_{\gamma+4}}E_0(k)+
\dfrac{3C^\gamma\mu}{2g_{\gamma+3}}E_1(k)+
$$
$$
+
C^\gamma|\mu|\Bigg[-(1+q)B^+|\mu|+(1-q)\varepsilon_TC-C^{2\gamma}\times
$$
$$
\times\dfrac{1-q}{\pi}\int\limits_{0}^{\infty}
\Bigg[\dfrac{CE_0(k_1)}{2g_{\gamma+4}}+
\dfrac{3\mu E_1(k_1)}{2g_{\gamma+3}}
\Bigg]\dfrac{dk_1}{C^{2\gamma}+k_1^2\mu^2}\Bigg].
\eqno{(13)}
$$

We consider equality (10).
Rewriting it in terms of the Fourier integrals, we obtain
$$
E_j(k)=$$$$+\int\limits_{-1}^{1}\int\limits_{0}^{\infty}
\mu^j\Phi(k,\mu,C)C^{\gamma+3}g(C)d\mu dC,\quad j=0,1.
\eqno{(14)}
$$

We solve Eq. (13) for $\Phi(k,\mu,C)$ and substitute
it in (14) for $j = 0,1$.
We obtain a system consisting of two characteristic equations.
Without writing them, we combine
them into one vector characteristic equation
$$
\Lambda(k)E(k)=-2(1+q)B^+T_1(k)+2(1-q)\varepsilon_T
T_2(k)-
$$
$$-
\dfrac{1-q}{\pi}\int\limits_{0}^{\infty}
J(k,k_1)E(k_1)dk_1.
\eqno{(15)}
$$

Here
$$
T_{m,l}(k)=\int\limits_{0}^{1}\int\limits_{0}^{\infty}
\dfrac{\mu^l C^mg(C)d\mu
dC}{C^{2\gamma}+k^2\mu^2},
$$
$\Lambda(k)$ is the dispersion matrix
$$ \extrarowheight=10pt
\Lambda(k)=\left [\begin{array}{cc}\extrarowheight=19pt
    1-\dfrac{1}{g_{\gamma+4}}T_{3\gamma+4,0}(k) &
    \dfrac{3ik}{g_{\gamma+3}}T_{2\gamma+3,2}(k) \\
    \dfrac{ik}{g_{\gamma+4}}T_{2\gamma+4,2}(k) &
    1-\dfrac{3}{g_{\gamma+3}}T_{3\gamma+3,2}(k) \\
  \end{array}
\right],
$$
$E(k)$ is an unknown column vector
$$ %\extrarowheight=10pt
E(k)=\left[\begin{array}{c}
           E_0(k) \\
           E_1(k) \\
         \end{array}
       \right],
$$%\extrarowheight=19pt
$T_1(k)$ and $T_2(k)$ are the column vectors of constant terms
$$
T_1(k)=\left[\begin{array}{c}
              T_{3\gamma+3,2}(k) \\
              -ikT_{2\gamma+3,4}(k) \\
            \end{array}
          \right],
$$
$$
T_2(k)=\left[\begin{array}{c}
              T_{3\gamma+4,1}(k) \\
              -ikT_{2\gamma+4,3}(k) \\
            \end{array}
          \right],
$$
and $J(k,k_1)$ is the matrix function
$$\extrarowheight=10pt
J(k,k_1)=\left [\begin{array}{cc}
    \dfrac{1}{g_{\gamma+4}}J_{4\gamma+4,1}(k,k_1) &
   -\dfrac{3ik}{g_{\gamma+3}}J_{3\gamma+3,3}(k,k_1) \\
    -\dfrac{ik}{g_{\gamma+4}}J_{3\gamma+4,3}(k,k_1) &
    \dfrac{3}{g_{\gamma+3}}J_{4\gamma+3,3}(k,k_1) \\
  \end{array}
\right]
$$
which we call the Neumann matrix.

The integrals
$$
J_{m,l}(k,k_1)=\int\limits_{0}^{1}\int\limits_{0}^{\infty}
\dfrac{\mu^l C^mg(C)d\mu
dC}{(C^{2\gamma}+k^2\mu^2)(C^{2\gamma}+k_1^2\mu^2)}
$$
are introduced in the Neumann matrix.

\section{5.  Method of successive approximations}

We seek the solution of Eq. (15) in the form
$$
\varepsilon_T=\dfrac{1+q}{1-q}\Big[\varepsilon_0+\varepsilon_1(1-q)+
\varepsilon_2(1-q)^2+\cdots\Big]
\eqno{(16)}
$$
and
$$
E(k)=2(1+q)\times $$$$ \times\Big[E^{(0)}(k)+E^{(1)}(k)(1-q)+E^{(2)}(k)(1-q)^2+\cdots\Big].
\eqno{(17)}
$$

We substitute these equalities in the characteristic equation.
We obtain the countable system of equations
$$
\Lambda(k)E^{(0)}(k)=-B^+T_1(k)+\varepsilon_0T_2(k),
\eqno{(18)}
$$
$$
\Lambda(k)E^{(1)}(k)=\varepsilon_1T_2(k)-\dfrac{1}{\pi}
\int\limits_{0}^{\infty} J(k,k_1)E_0(k_1)dk_1,
\eqno{(19)}
$$
$$
\Lambda(k)E^{(2)}(k)=$$$$+\varepsilon_2T_2(k)-\dfrac{1}{\pi}
\int\limits_{0}^{\infty} J(k,k_1)E_1(k_1)dk_1,\cdots.
\eqno{(20)}
$$

We transform the elements on the
principal diagonal of the dispersion matrix:
$$
1-\dfrac{1}{g_{\gamma+4}}T_{3\gamma+4,0}(k)=
\dfrac{k^2}{g_{\gamma+4}}T_{\gamma+4,2}(k),
$$
$$
1-\dfrac{1}{g_{\gamma+3}}T_{3\gamma+4,2}(k)=
\dfrac{3k^2}{g_{\gamma+3}}T_{\gamma+3,4}(k).
$$

The dispersion matrix function now becomes
$$
\Lambda(k)=\left [\begin{array}{cc}
  \dfrac{k^2}{g_{\gamma+4}}T_{\gamma+4,2}(k)&
  \dfrac{3ik}{g_{\gamma+3}}T_{2\gamma+3,2}(k) \\
    \dfrac{ik}{g_{\gamma+4}}T_{2\gamma+4,2}(k) &
    \dfrac{3k^2}{g_{\gamma+3}}T_{\gamma+3,4}(k) \\
  \end{array}
\right].
$$

We call the determinant of the dispersion matrix the dispersion function:
$$
\lambda(z)\equiv\det \Lambda(z)=k^2\omega(k),
$$
where
$$
\omega(k)=\dfrac{3k^2}{g_{\gamma+3}g_{\gamma+4}}\times $$$$ \times
\Big[k^2T_{\gamma+4,2}(k)T_{\gamma+3,4}(k)+
T_{2\gamma+3,2}(k)T_{2\gamma+4,2}(k)\Big].
$$

The matrix inverse to the dispersion matrix is
$$
\Lambda^{-1}(k)=\dfrac{D(k)}{\lambda(k)},
$$,
where
$$\extrarowheight=10pt
D(k)=\left [  \begin{array}{cc}
\dfrac{3k^2}{g_{n+3}}T_{n+3,4}(k)
 & -\dfrac{3ik}{g_{n+3}}T_{2n+3,2}(k) \\
   - \dfrac{ik}{g_{n+4}}T_{2n+4,2}(k) &
   \dfrac{k^2}{g_{n+4}}T_{n+4,2}(k)   \\
  \end{array}
\right ].
$$

We consider the construction of series (17). We set
$$ %\extrarowheight=10pt
E^{(0)}(k)=\left[\begin{array}{c}
           E_0^{(0)}(k) \\
           E_1^{(0)}(k) \\
         \end{array}
       \right].
$$%\extrarowheight=19pt

We consider the construction of series (17). We set
$$
E^{(0)}(k)=2\dfrac{D(k)}{\lambda(k)} \times $$$$\times
\left[
  \begin{array}{c}
    -B^+T_{3\gamma+3,2}(k)+\varepsilon_0T_{3\gamma+4,1}(k)  \\
    ik(B^+T_{2\gamma+3,4}(k)+\varepsilon_0T_{2\gamma+4,3}(k))  \\
  \end{array}
\right].
\eqno{(21)}
$$

Substituting expression (21) in Eq. (19), we find $E^{(1)}(k)$.
Substituting $E^{(1)}(k)$ in Eq. (20), we then find $E^{(2)}(k)$.
Continuing this process
without bound, we construct all terms of series (17).

We now show how to construct the terms of series (16).
For this, we write vector equality (21)
in the form of two scalar equalities:
$$
E_0^{(0)}=\dfrac{1}{k^2\omega(k)}
\Bigg\{-\dfrac{3k^2}{g_{\gamma+3}}T_{\gamma+3,3}(k) \times
$$$$\times
\Big[B^+T_{3\gamma+3,2}(k)-\varepsilon_0T_{3\gamma+4,1}(k)\Big]+
$$
$$
+
\dfrac{3k^2}{g_{\gamma+3}}T_{2\gamma+3,2}(k)%\times $$$$\times
\Big[B^+T_{2\gamma+3,4}(k)+\varepsilon_0T_{2\gamma+4,3}(k)\Big]
\Bigg\}
\eqno{(22)}
$$
$$
E_1^{(0)}=\dfrac{1}{k^2\omega(k)}\times $$$$\times
\Big\{ikT_{2\gamma+4,2}(k)
\Big[B^+T_{3\gamma+3,2}(k)-\varepsilon_0T_{3\gamma+4,1}(k)\Big]+
$$
$$
+
ik^3T_{\gamma+4,2}(k)\Big[B^+T_{2\gamma+3,4}(k)+
\varepsilon_0T_{2\gamma+4,3}(k)\big]\Big\}.
\eqno{(23)}
$$

It can be seen from expression (22) that the function $E_0^{(0)}(k)$
has no singularities at zero, while the function $E_1^{(0)}(k)$
according to (23) has a second--order pole at $k=0$.
Eliminating the singularity at zero, we obtain
$$
\varepsilon_0=B^+\dfrac{T_{3\gamma+3,2}(0)}{T_{3\gamma+4,1}(0)}=
B^+\dfrac{2g_{\gamma+3}}{3g_{\gamma+4}}.
$$

The quantity $B^+$ is proportional to the heat flux:
$$
\mathbf{Q}=\int f(x,\mathbf{p})
\dfrac{\partial \varepsilon (p)}{\partial \mathbf{p}}
\,\varepsilon(p)d\Omega_B.
$$

We transform this expression as
$$
\mathbf{Q}=\dfrac{(2s+1)u_0^2}{(2\pi \hbar)^3}
\int h(x,\mathbf{p})g(p)\mathbf{p}\,d^3p.
$$

Integrating in this expression over the dimensionless momentum, we obtain
$$
\mathbf{Q}=\dfrac{(2s+1)(kT_s)^4}{(2\pi \hbar)^3u_0^2}
\int h(x,\mathbf{C})g(C)\mathbf{C}\,d^3C.
$$

We replace here  the function $h(x,\mathbf{C})$ with its Chapman --- Enskog
expansion $h_{as}(x,\mathbf{C})$. As a result, we have
$$
Q_x=\dfrac{(2s+1)(kT_s)^4}{(2\pi \hbar)^3u_0^2}
\int\limits_{-1}^{1}
\int\limits_{0}^{\infty}\int\limits_{0}^{2\pi}
\Big[\varepsilon_TC+B^+\mu\Big]\times $$$$ \times
g(C)C^3\mu d\mu dC
d\chi=%$$$$=
\dfrac{(2s+1)(kT_s)^4}{(2\pi \hbar)^3u_0^2}\cdot
\dfrac{4\pi}{3}g_3B^+
$$
for the $x$ component of the heat flux.

We hence have
$$
B^+=\dfrac{6\pi^2}{g_3}\cdot\dfrac{\hbar^3u_0^2Q_x}{(2s+1)(kT_s)^4}.
$$

The quantity $\varepsilon_0$ is therefore
$$
\varepsilon_0=\dfrac{4\pi^2g_{\gamma+3}}{g_3 g_{\gamma+4}}
\cdot \dfrac{\hbar^3 u_0^2}{(2s+1)(kT_s)^4}Q_x.
\eqno{(24)}
$$

Returning to the formula for the temperature jump
$$
\Delta T=RQ_x,
$$
we find from expression (24) that the
Kapitsa resistance in the zeroth approximation is
$$
R=\dfrac{4\pi^2g_{\gamma+3}}{g_3g_{\gamma+4}}\cdot\dfrac{1+q}{1-q}
\cdot\dfrac{\hbar^3 u_0^2}{(2s+1)k^4T_s^3}.
$$

\begin{figure}[t]
\begin{center}
\includegraphics[width=8cm, height=7cm]{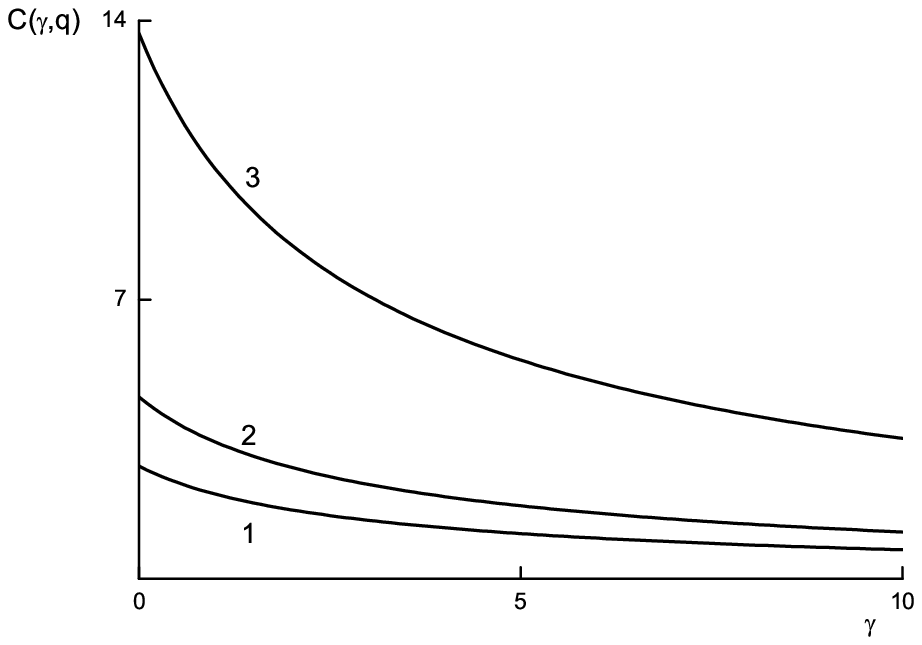}
\end{center}
\begin{center}
Fig. 1.   Dependence of the temperature jump coefficient
on the parameter $\gamma$: the specular reflection coefficient is
$q = 0.3$ for curve $1$,
$q = 0.5$ for curve $2$, and $q = 0.8$ for curve $3$.
\end{center}
\begin{center}
\includegraphics[width=8cm, height=7cm]{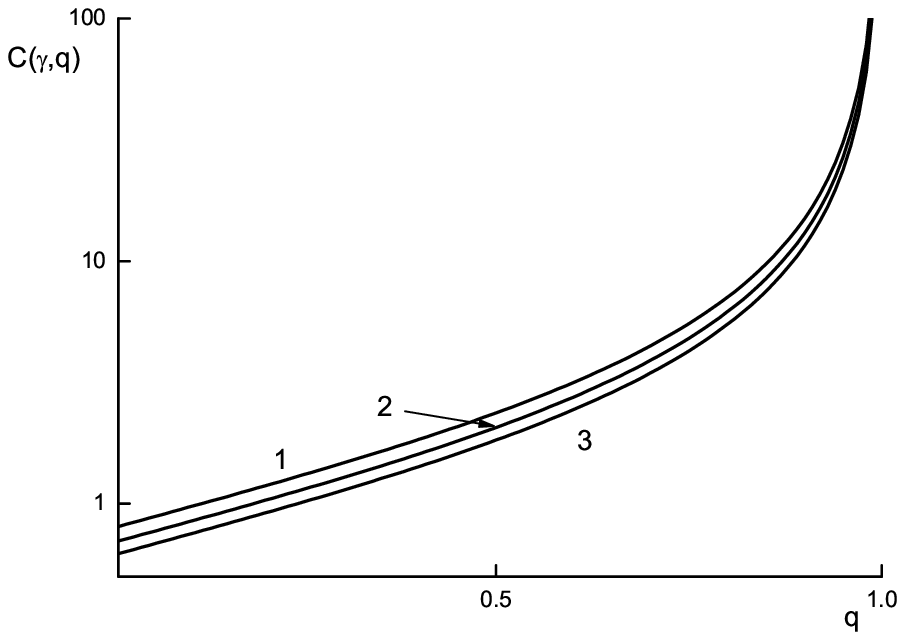}
\end{center}
\begin{center}
Fig. 2.   Dependence of the temperature jump coefficient
on the specular reflection coefficient q: the collision
parameter is $\gamma = 3$ for
curve $1$, $\gamma=4$ for curve $2$, and $\gamma=5$ for curve $3$.
\end{center}
\end{figure}

We rewrite this equation in the form
$$
R=C(\gamma,q)\cdot\,\dfrac{\hbar^3 u_0^2}{(2s+1)k^4T_s^3},
$$
where
$$
C(\gamma,q)=\dfrac{4\pi^2 g_{\gamma+3}}{g_3g_{\gamma+4}}\cdot
\dfrac{1+q}{1-q}
\eqno{(25)}
$$
is the (dimensionless) coefficient of the
temperature jump.

The plus sign in formula (25)
indicates that the wall temperature
is higher that the phonon component temperature.

The graphs of the behavior of the temperature jump
coefficient as a function of the parameter $\gamma$ and
the specular reflection coefficient $q$ are shown
in Figs. 1 and 2.

It can be seen from these graphs
that the quantity $C(\gamma,q)$ decreases monotonically
as the parameter $\gamma$ increases. As the specular reflection
coefficient $q$ tends to unity, the quantity $C(\gamma,q)$
increases without bound because the heat exchange between
the wall and the
gas adjoining it becomes impossible in this limit.

\section{6. Conclusions}

We have constructed a kinetic equation for a degenerate
quantum Bose gas whose collision rate depends on the
momentum of elementary excitations of the Bose gas.
We considered the case where the phonon component is
the key factor in the elementary excitations. The
boundary conditions were assumed to be specular--diffuse.
We solved the half--space boundary value problem of the
temperature jump at the boundary of a degenerate gas in
the presence of a Bose --- Einstein condensate. We derived
a formula for finding the temperature jump and calculating
the Kapitsa resistance.

We have developed a sufficiently general method for solving
kinetic equations with specular--diffusive boundary conditions;
this method was first proposed in \cite{Lat6}, where the problem of
the skin effect was considered.

\end{document}